%% file: main.tex
\documentclass[12pt,a4paper]{spie}  
\usepackage[english]{babel}   
\usepackage{amsmath,amsfonts,amssymb}
\usepackage{textcomp,gensymb}
\usepackage{graphicx}
\usepackage{epstopdf}
\usepackage[separate-uncertainty=true,multi-part-units=single]{siunitx}
 \DeclareSIUnit\bar{bar}
\usepackage{float}

\usepackage{placeins}
\usepackage{subcaption}

\usepackage{booktabs,dcolumn}

\usepackage[colorlinks=true, allcolors=blue]{hyperref}
\title{Second-Generation Fabry-Pérot Unit for CARMENES}

\author[a]{Hanna Lina Pleteit}
\author[a,*]{Michael Debus}
\author[a]{Sebastian Schäfer}
\author[a]{Ansgar Reiners}
\affil[a]{Institut f\"ur Astrophysik und Geophysik, Georg-August-Universit\"at, Friedrich-Hund-Platz 1, G\"ottingen, Germany}

\authorinfo{{*} Send correspondence to M.D.: mdebus@phys.uni-goettingen.de }

\graphicspath{{plots/}} 

\begin{document} 
\maketitle

\input{sections/00-abstract.tex}


\input{sections/01-intro}
\input{sections/02-simulations}

\input{sections/03-design}

\input{sections/04-labtest}

\input{sections/05-integration}
\input{sections/06-summary}

\subsection* {Acknowledgments}  
We thank Calar Alto Observatory staff for their kind help during the integration of the new etalon system. 

\bibliography{CARFP.bib} 
\bibliographystyle{spiebib} 
\end{document}

%% file: sections/00-abstract.tex
\begin{abstract}
CARMENES consists of two high-resolution spectrographs located at the Calar Alto \SI{3.5}{\meter} telescope, covering the wavelength range from \SIrange{520}{1710}{\nano\meter}. The original Fabry-Pérot (FP) units have been essential to track and remove instrumental drift while also improving the wavelength solution provided by the hollow cathode lamps. In this paper we present the second-generation FP unit that has begun operation at Calar Alto in April 2024.
It features drastically improved temperature stabilization, additional temperature monitoring, and enhanced optomechanics. Temperature stabilization is improved through better vacuum vessel design, thermal shielding and improved thermal control. The temperature is now monitored with Pt-100 sensors inside the vacuum vessel, one of which is used to drive the temperature control. The new purpose-built optomechanics allow for an improved alignment procedure leading to higher line contrast. The etalon coatings now cover a broader wavelength range, the optics have been upgraded, and the fibers have been updated to an octagonal shape. 
In laboratory measurements with a Fourier transform spectrometer we compare the FP radial velocity (RV) drift to that of an iodine cell. In a two hour binning we achieve an RMS of the differential RVs of \SI{6}{\centi\meter\per\second}. The new FP unit has already been successfully integrated at CARMENES, with first calibration data indicating improved RV precision.
\end{abstract}

\keywords{CARMENES, Fabry-Pérot etalon, spectrograph calibration, high resolution spectroscopy, radial velocity method}

%% file: sections/01-intro.tex
\section{Introduction} \label{sec:intro}

The Calar Alto high-Resolution search for M dwarfs with Exo-earths with Near-infrared and optical Echelle Spectrographs (CARMENES) survey uses high-resolution radial velocity measurements to detect low-mass planets around M dwarfs. The original CARMENES survey has detected a total of 22 new planets between 2016 and 2020. It further proved its capabilities for the follow-up of transiting exoplanets and spectroscopy of planetary atmospheres\cite{quirrenbach_carmenes_2020}. Building on this success the CARMENES Legacy-Plus survey targets another 387 M dwarfs until the end of 2026. Additionally, the CARMENES spectrographs are and continue to be one of the major instruments at the Calar Alto Observatory open to observation time proposals from Spanish institutions or collaborations therewith. Thus a number of small-scale instrumental upgrades on the spectrograph and calibration unit are scheduled. Our focus is on the calibration unit. The RV precision on the order of \SI{1}{\meter\per\second} required for low-mass targets requires precise monitoring of the instrumental drift.\cite{quirrenbach_carmenes_2012} To this end CARMENES employs a combination of hollow cathode lamps and Fabry-Pérot etalons (FPs). The wavelength solution is generated by measuring both light sources at the beginning and the end of the night. Additionally, to track the instrumental drift during the night, the FP is measured simultaneously with the science light in a second fiber\cite{quirrenbach_carmenes_2018}.

The main advantage of the FPs is that they provide a robust, stable calibration light source with a large number of densely spaced, similarly bright lines over the extensive wavelength region covered by CARMENES from \SIrange{550}{1710}{\nano\meter}. Additionally, the line width and spacing can be easily specified to the requirements of the spectrograph as they are functions of the mirror distance and reflectivity of the FP cavities only. Current state-of-the-art spectrographs like ESPRESSO\cite{schmidt_fundamental_2021}, HPF\cite{jennings_frequency_2020} and SPIRou\cite{cersullo_new_2017} use them to reach RV precision below \SI{1}{\meter\per\second}. While not providing an accurate wavelength solution, they have multiple advantages over laser frequency combs. They are more reliable and provide a better intensity stability in addition to being available at a significantly lower cost.

In this work we present a design that is an evolution of the first generation CARMENES Fabry-Pérot unit \cite{schafer_two_2018} which has been in operation since 2016. Most notably the temperature stabilization is revised. This includes an improved vacuum vessel design, further thermal shielding and temperature monitoring by six Pt-100 sensors at different locations inside the vacuum vessel. One of the sensors inside the vessel will be used for controlling the temperature instead of the thermostat read-out. Following concerns about the alignment of the first generation FPs, we used simulations to verify that the FPs are indeed misaligned. Consequently, we employ an improved alignment procedure. This should lead to better contrast in the calibration spectra.

The updated vacuum vessel builds on the design by Debus et al. from 2022\cite{debus_near-infrared_2022}. That FP setup was built for the Fourier-Transform-Spectrograph (FTS) at the IAG. The ultra high resolution of the FTS (R = 700 000) allowed to measure the drift of the FP against an iodine cell reference with high precision. Over 90 hours no significant long term trend was observed with an RMS of the RV differences of \SI{10.7}{\centi\meter\per\second} close to the average uncertainty of \SI{10.2}{\centi\meter\per\second} in 10 minute measurements\cite{debus_towards_2023}. With the second generation CARMENES Fabry-Pérot unit we aim to achieve the same RV stability and thus improve the calibration of the CARMENES spectrograph.

This paper is organized as follows. We start by looking at simulation results for the FPs in Section \ref{sec:simulations} that lead to the optical part of the design described in Section \ref{sec:setup}. In Section \ref{sec:measurements} we document the performance of the whole Fabry-Pérot unit with respect to temperature and radial velocity (RV) stability. Section \ref{sec:conc} concludes with a brief summary and discussion followed by an outlook.

%% file: sections/02-simulations.tex
\section{Simulations} \label{sec:simulations}
To understand the effects of various design choices in optics, such as finesse, cavity width, fiber size, and collimation focal length, as well as the effects of misalignment on the final measured spectrum of the Fabry-Pérot (FP) interferometer, we simulated the FP spectrum. This approach is similar to that taken by Schäfer\cite{schafer_fabry-perot_2014}.

For the simulations, a forward model of the FP was constructed using the Fabry-Pérot equation\cite{ade_physical_2022}:
\begin{equation}
    \label{equ:FPintensity}
    I(\lambda) = \frac{I_0}{1+F\sin^2{(\frac{2\pi}{\lambda}nl\cos{(\theta)})}} 
\end{equation}
with the intensity $I_0$, the cavity length $l$, the angle of incidence $\theta$ and the coefficient of finesse 
\begin{equation}
    F \approx \left(\frac{2\mathcal{F}}{\pi}\right)^2.
\end{equation}
As a starting point we use the parameters of the previous CARMENES FP unit: a cavity length of $l=\SI{9.99}{mm}$ for the VIS FP or $l = \SI{12.334}{mm}$ for the NIR FP. Both etalons have an average mirror reflectance of 69\,\% which amounts to a finesse value of $\mathcal{F}=8.418$. Because the etalons are kept inside a vacuum chamber we can assume the index of refraction $n_{vac}=1$. In addition to the FP, on the input side the fiber and the light source are also modeled. A $\SI{50}{\micro m}$ fiber acts as an extended light source, meaning that a distribution of different ray angles is entering the FP. To model this, we use a set of 25 concentric annuli, for which the intensities are calculated according to Equation\,\eqref{equ:FPintensity} and then summed up, weighted by their area as we assume a homogeneous light distribution. The light source is modeled as a black body with the lamp temperature ($T=\SI{2792}{K}$) and multiplied onto the spectrum. 
On the output side to model the detector, the spectrum is folded with a Gaussian corresponding to the resolving power of $R =94,600$ for the visible spectrograph and $R = 80,400$ for the near-infrared. Afterwards the pixel sampling is added by integrating over the wavelengths falling into a single pixel. Since the pixel sampling varies non-linearly over the detector, we take the measured values from CARMENES for the wavelength ranges used. For the visible that we evaluate the model at \SI{795}{nm} with a pixel sampling of 2.49 and for the near infrared at \SI{1200}{nm} with a pixel sampling of 2.69. As the final step, the spectrum is normalized.

\begin{figure}[b]
\centering
 \includegraphics[width=\textwidth]{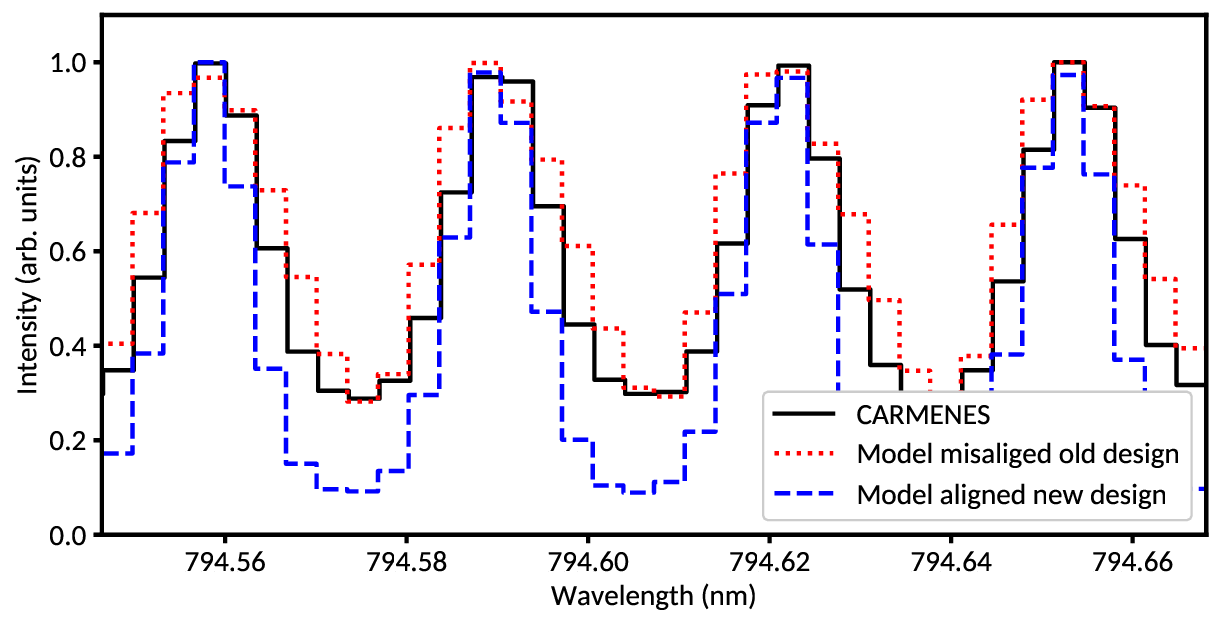}
 \caption{Comparison of the CARMENES spectrum of the first generation VIS FP (Black solid line) with the forward model of the current setup (red dotted line). The incident ray angle in the forward model is chosen to be $\theta = \SI{39}{mrad}$ so that the contrast matches with the CARMENES measurement. The forward model of an improved FP  setup is also shown (blue dashed line). The alignment is assumed to be perfect ($\theta = \SI{0}{mrad}$) and a collimator with a longer focal length of $f=\SI{50.8}{mm}$ is used.} \label{fig:oldnewvismodelvscar}
\end{figure}

Fig.\,\ref{fig:oldnewvismodelvscar} shows a modeled spectrum of the first generation CARMENES VIS FP. One important characteristic of the spectrum is the contrast, which is the relative difference between the peak and the background intensity in between the peaks. A high contrast in the Fabry-Pérot calibration spectra leads to a higher signal-to-noise-ratio which improves the precision with which the FP peaks can be determined. In the comparison between the modeled and the measured VIS FP, the contrast serves as an indicator to determine the misalignment of the FP. Since the incident ray angle is a parameter of our model, it can be tuned until the model and the data match with respect to the contrast, as shown in Fig.\,\ref{fig:oldnewvismodelvscar}. The resulting misalignment angle of $\theta = \SI{39}{mrad}$ is bigger than expected. Furthermore, the peaks in the modeled spectrum are broader than the ones in the measurement, indicating that a misalignment is not the only effect causing the broadening of the peaks. Likely, other factors we do not model, e.g. defocus, also play a role. This hints at problems with the alignment of the FP that could be improved by the second generation. Fig.\,\ref{fig:oldnewvismodelvscar} shows how the resulting spectrum could be improved according to the simulations: The forward model is perfectly aligned and for the collimating element between the input fiber and the FP, a longer focal length of $f=\SI{50.8}{mm}$ compared to the $\SI{33}{mm}$ in the current setup is chosen. This is advantageous, because a collimator with a longer focal length means that the light distribution of the extended light source translates to smaller deviations in the collimated beam on the input of the FP. Fig.\,\ref{fig:oldnewvismodelvscar} shows that consequently the contrast improves significantly and the lines become narrower as well.



\begin{figure}[htb]
\centering
\begin{subfigure}{0.49\textwidth}
 \includegraphics[width=\textwidth]{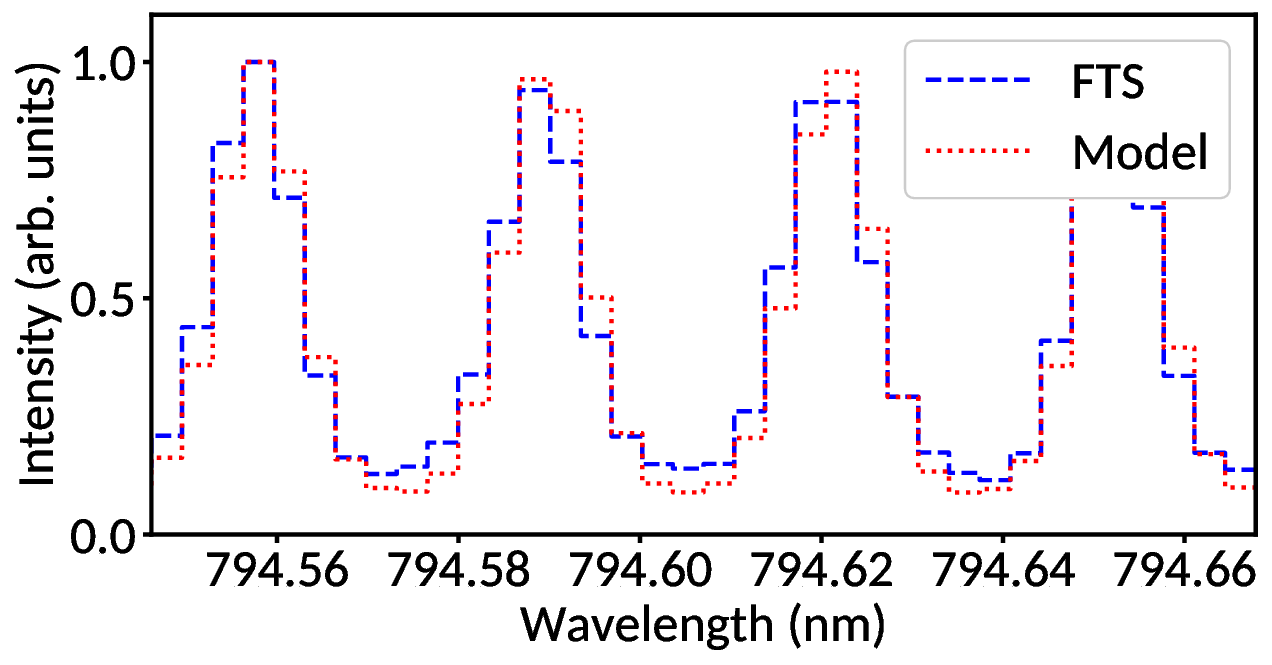}
 \caption{VIS FP} \label{fig:vismodelvsfts}
\end{subfigure}
 \begin{subfigure}{0.49\textwidth}
 \includegraphics[width=\textwidth]{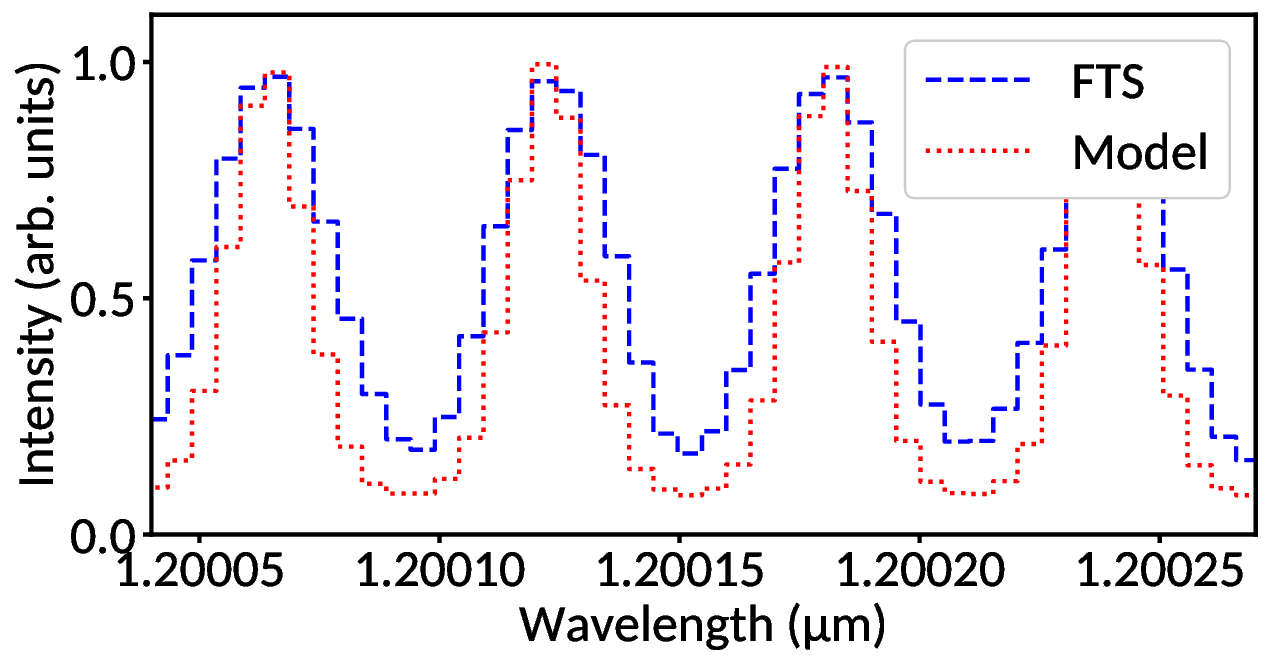}
 \caption{NIR FP.} \label{fig:nirmodelvsfts}
\end{subfigure}
\caption{Comparison of the aligned FPs as measured by the FTS and folded with the CARMENES resolving power (VIS =\SI{94,600}{} ; NIR = \SI{80,400}{}) and sampled with the corresponding pixel sampling (blue dashed line) (VIS 2.49; NIR 2.69)  with the forward modeled spectrum (red solid line).}
\end{figure}

After designing the new FPs with the desired specifications (see Sec.\,\ref{sec:setup}), we can measure how close the alignment in our lab comes to achieving the optimum given by the forward model. Since our lab is equipped with an FTS with a much higher resolution than the CARMENES spectrographs, the measured spectra are folded with the CARMENES resolution and binned according to the pixel sampling for the comparison. Fig.\,\ref{fig:vismodelvsfts} shows the comparison between the forward model and the measurement of what we ascertained to be the best alignment, as measured by the alignment method developed by Debus et al.\cite{debus_near-infrared_2022} for the VIS FP . The contrast of the aligned FP comes very close to the goal given by the model. The differences can be accounted for by the fact that we did not model the effects of defective or non-parallel mirrors, both of which depreciate the finesse. For the NIR FP the same comparison is done. Fig.\,\ref{fig:nirmodelvsfts} shows the alignment measurement in our lab, which in this case is slightly further away from the modeled optimum.

%% file: sections/03-design.tex
\section{Design} \label{sec:setup}

The Fabry-Pérot unit, shown in Fig.\,\ref{fig:temperaturesensors}, consists of four distinct parts. Firstly, the Fabry-Pérot etalons are at the heart of the design. They sit on custom-made optical benches that provide all the alignment equipment and optical elements around the FPs. This optical setup, in turn, is housed inside the vacuum vessel with its temperature stabilization and monitoring. The last part comprises the light sources and fibers that bring light into the setup (not shown in the drawing).

\subsection{Fabry-Pérot etalon}
\begin{figure}[htb]
   \centering
\begin{subfigure}{0.49\textwidth}
\includegraphics[width=\textwidth]{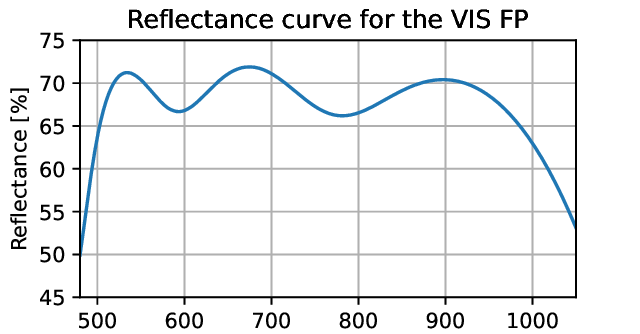}
\caption{VIS}\label{fig:reflectancevis}
\end{subfigure}
\begin{subfigure}{0.49\textwidth}
\includegraphics[width=\textwidth]{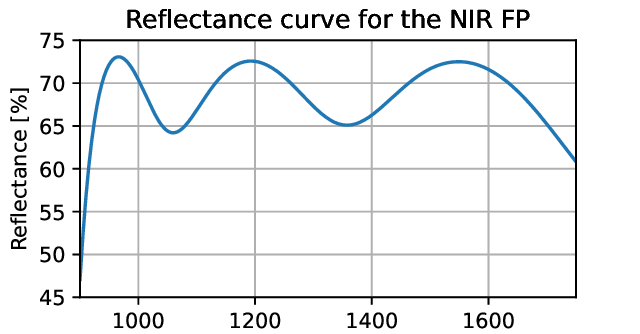}
\caption{NIR}\label{fig:reflectancenir}
\end{subfigure}
   \caption{Specified coating reflectance for the two FPs. The average reflectance is 69 \% for both.}
   \label{fig:reflectance}
 \end{figure}

For the Fabry-Pérot etalons we chose air-gap etalons from SLS Optics. The two etalons are soft-coated to the same finesse of $\mathcal{F} \approx 8$. Otherwise, they differ in their properties to serve the corresponding spectrograph. The optical properties are the same as in the first generation of CARMENES etalons \cite{schafer_two_2018}. Only the spectral coverage was extended and the clear aperture enlarged to allow for collimators with a longer focal length that inevitably produce a larger beam diameter at the position of the etalon. The reflectance as a function of wavelength, as specified by the manufacturer, can be found in Fig.\,\ref{fig:reflectance}. All specifications for the etalons can be found in Tab.\,\ref{tab:FPSpecs}. 

\begin{table}[htbp]
    \centering
    \begin{tabular}{lcc}
        \toprule
        \textbf{Specification} & \textbf{VIS FP} & \textbf{NIR FP} \\
        \midrule
        Spectral Coverage [nm] & 500-1000 & 950-1710\\
        Clear aperture [mm] & 30 & 30\\
        Mirror distance [mm] & 10.00 &12.33\\
        Mirror reflectance avg. [\%] & 69& 69 \\
        Finesse & $\approx 8.4$ & $\approx 8.4$ \\
        FSR [GHz] & 15.0 & 12.1\\
        \bottomrule
    \end{tabular}
    \vspace{0.3cm}
    \caption{Specifications of the two Fabry-Pérot etalons vacuum-gap FP etalon from SLS Optics (UK).}
    \label{tab:FPSpecs}
\end{table}

\subsection{Optical benches}

\begin{figure}[htb]
\centering
 \includegraphics[width=0.7\textwidth,angle =180]{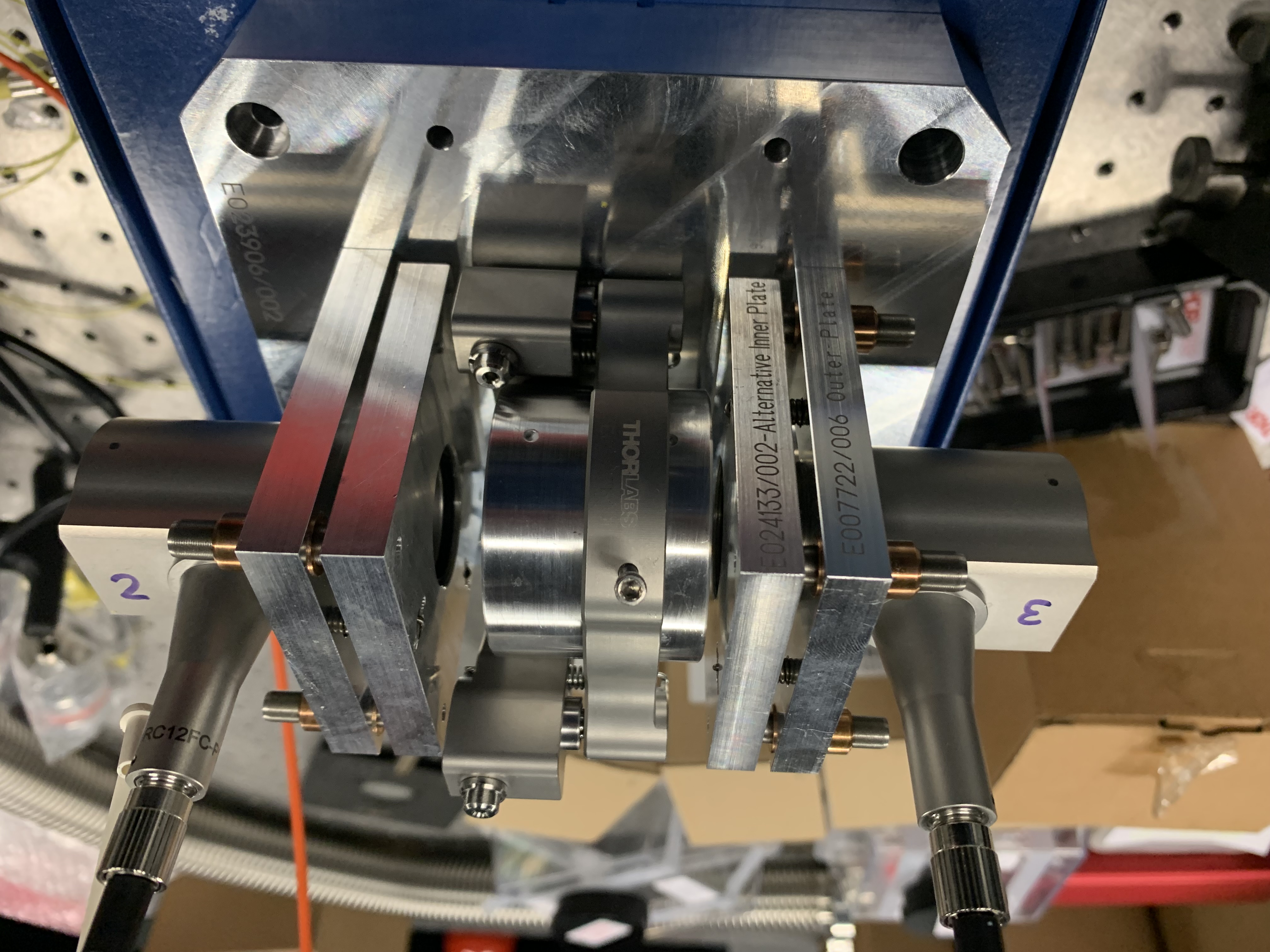}
 \caption{The VIS etalon on its optical bench, mounted in a POLARIS K2Vs2L mount and the custom made tip tilt stages for the RC12FC-P01 OAPs.} \label{fig:benches}
\end{figure}

As two FPs are housed in one vacuum chamber, the optical benches (see Fig.\,\ref{fig:benches}) have to be stacked in order to have enough room for both FPs. This is accomplished by a two-story design with identical optical setups, connected by aluminum walls that also establish thermal contact. The optical benches are chosen to be massive aluminum blocks to provide a high thermal mass with only minor differences. The upper bench has recesses for the fibers to reach the lower bench. The lower bench has venting grooves because it is directly connected to the bottom of the vacuum chamber. This was chosen after Debus et al.\cite{debus_towards_2023} showed that the bottom is the thermally most stable part of the vacuum chamber and has the additional benefit of dispensing with any complicated structures to keep the optical benches isolated from the vacuum vessel.

The optical setup on both benches is identical, albeit rotated by 90°. Vacuum compatible POLARIS K2VS2L (Thorlabs) stages are used to mount the etalons. The vertical mount saves space by keeping the adjustment screws out of the long axis of the setup. Light is coupled into and out of the etalon via two RC12FC-P01 (Thorlabs) off-axis parabolic mirrors connected to custom-made tip tilt stages that offer improved stability during transport. As the RC12FC-P01 fiber couplers suffer from a chronic defocus (probably due to problems in Thorlabs manufacturing process), we assessed the collimation quality with a shearing interferometer. We then manufactured shimming rings of the required thickness (ranging from \SIrange{0.2}{0.4}{\milli\meter} to place in the FC/PC coupler. With these we drastically improved the collimation quality.

\subsection{Vacuum vessel}

\begin{figure}\centering
    \includegraphics[height=0.9\textheight]{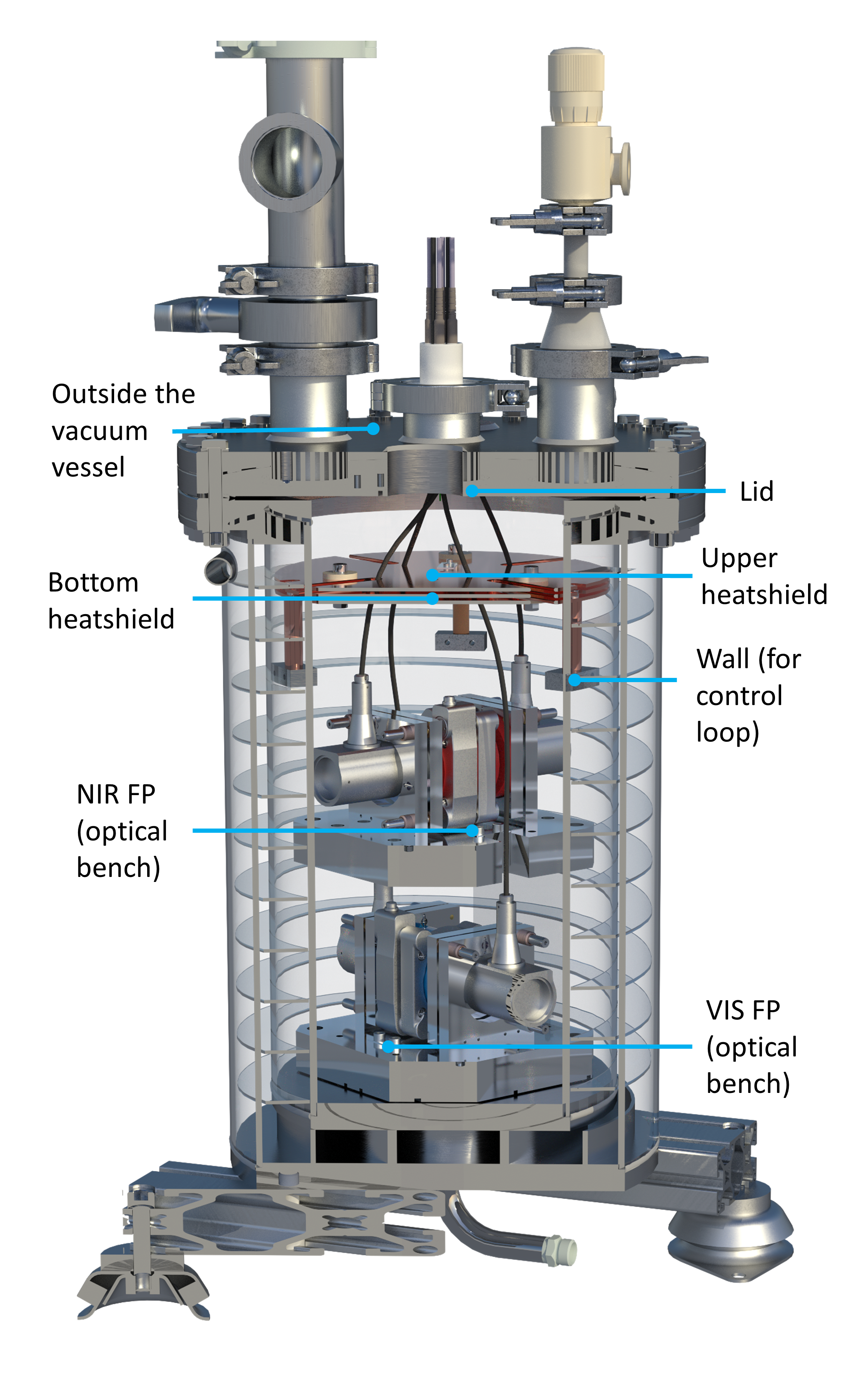}
 \caption{CAD model of the second generation CARMENES vacuum vessel. The dots show the positions of the six temperature sensors inside the vacuum vessel. The wall sensor is used for the temperature control loop and thus cannot be readout separately.}\label{fig:temperaturesensors}
\end{figure}

The diameter of the vacuum chamber is given by the DN 250 CF flange used as the lid. It is smaller than the first generation because several measures were taken to minimize the length of the optical setup. These include the vertical etalon mount and the custom-made tip tilt stages.
The thermal stabilization mechanism relies on silicon oil that is kept at a constant \SI{18}{\degree C} via a Huber Ministat 125 and flows through the double-walled walls and bottom of the vacuum vessel. A spiral structure (see Fig.\,\ref{fig:temperaturesensors}) guides the oil to ensure a continuous flow and even heat distribution. As the lid is the only part of the vacuum chamber that is not thermally stabilized an additional assembly of three heat shields is used to reduce the impact of changing room temperature. This is also an improvement to the one heat shield in the previous design. The copper heat shields are thermally insulated from each other by Macor washers. The lowest heatshield is fastened to three heat shield rests that are welded onto the walls and thus in good thermal contact with the stabilized part of the vessel. The high thermal conductivity of the copper results in a homogeneous temperature distribution with the three heat shields forming a temperature gradient from the non-insulated lid to the inner part of the vacuum chamber.
The lid provides 5 KF 40 flanges for the HiCube80Eco vacuum pump, the HPT 200 pressure gauge, the fiber feedthrough, a pressure relief valve and the electric feedthrough (Pavemate 1637-NW40, according to MIL-C 26482 standard) for six PT-100 Class A temperature sensors. Those are distributed throughout the vacuum chamber to monitor the temperature. Their positions can be found in Fig.\,\ref{fig:temperaturesensors}. The sensors are read out using two OMEGA PT-104A RTD input DAQ modules.

\subsection{Styrodur enclosure}
To further improve the thermal stability, the vacuum vessel is placed inside a styrodur enclosure. The vessel itself is fixed to a wooden baseplate using machine feet which provide some mechanical dampening. On the wooden plate a frame of aluminum profiles is fixed in which a lifting mechanism for the lid is integrated. Using nylon screws, two layers of \SI{40}{mm} styrodur plates are attached to the aluminum profiles. To prevent convection at the edges between different plates, a step profile is used at the edges. Fitting exit holes for electric, vacuum, thermal and fiber connections are provided.

\subsection{Light source and fibres}
The light for the calibration spectrum is generated by two Thorlabs SLS201L/M halogen lamps that produce less heat than the current light sources for the first generation. To couple the light into the setup four fibers are used in conjunction with a fiber feedthrough flange of the vacuum chamber. Two \SI{50}{\micro\meter} input fibres and two \SI{89}{\micro\meter} output fibers from CerampOptec with an octagonal core are chosen as they provide better scrambling. All fiber connectors at the light sources and OAPs are FC/PC standard selected for their superior centering.

%% file: sections/04-labtest.tex
\section{Laboratory Testing} \label{sec:measurements}
To verify the performance of the FP setup we extensively tested it in our laboratory using the internal Pt-100 temperature sensors inside the FP vacuum vessel as well as measurements with the FTS. With the FTS we recorded master spectra using the lamps also employed at the observatory for both the NIR and VIS FP to characterize their wavelength-dependent effective gap size as well as RV-drift measurements (with the VIS FP only) with LEDs for higher flux and an iodine cell as an absolute drift reference, similar to the setup described in \cite{debus_towards_2023}. All FTS spectra were taken with the FTS evacuated to \SI{0.2}{\milli\bar} at a resolution of \SI{0.02}{\per\centi\meter} corresponding to a resolving power of \SI{385000}{} at \SI{1400}{\nano\meter} and \SI{715000} at \SI{700}{\nano\meter}.

\subsection{RV-drift during thermalization}
\begin{figure}
    \centering
   \includegraphics[width=0.9\textwidth]{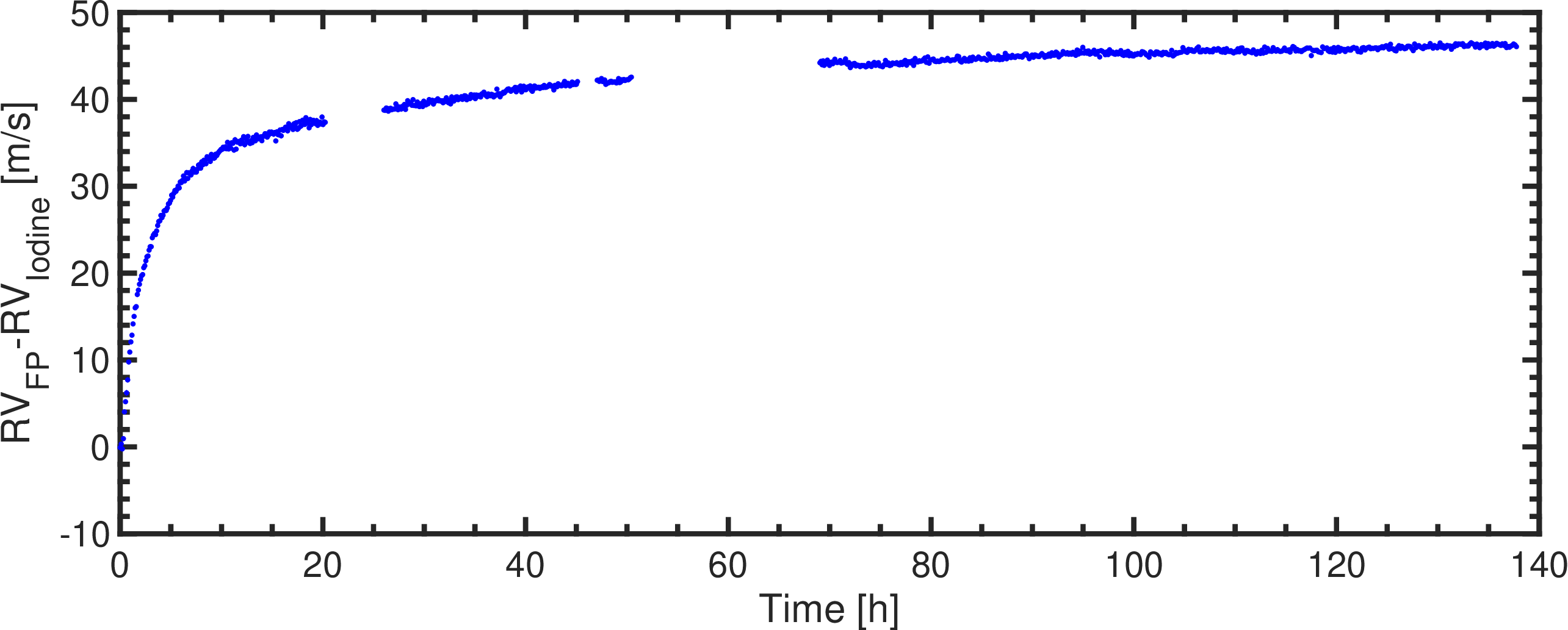}
    \caption{The RV drift of the FP after starting the temperature control with a setpoint \SI{3.3}{\kelvin} below room temperature measured against an iodine cell to take out the instrumental drift of the FTS.}
    \label{fig:rvsettling}
\end{figure}

To get a better understanding of the magnitude of temperature induced RV drifts we tracked the RV drift with the FTS while we turned on the temperature control for the first time. We set the temperature to \SI{18}{\celsius} as this is the temperature set point that will be used at the Calar Alto observatory. This was offset from the laboratory room temperature by about \SI{3}{\kelvin}. The RV-drift induced by this temperature shift is shown in Fig.\,\ref{fig:rvsettling}.  At the optical benches, the overall  temperature step was about \SI{3.3}{\kelvin}, while the measured overall RV drift was \SI{46}{\meter\per\second}. Using the thermal expansion coefficient of zerodur of about \SI{6}{\meter\per\second\kelvin} we can estimate the RV drift induced by the shrinking of the zerodur spacer to about \SI{20}{\meter\per\second}. The remaining RV drift of about \SI{26}{\meter\per\second} is most likely caused by misalignment and defocussing due to the temperature shift. This component therefore amounts to about \SI{0.8}{\centi\meter\per\milli\kelvin} in our setup for a total of \SI{1.4}{\centi\meter\per\milli\kelvin}, which indicates a temperature stability better than \SI{7.1}{\milli\kelvin} is needed to keep the RV drift below \SI{10}{\centi\meter\per\second}. This rough calculation however is not taking the time it takes for the RVs to follow the temperature into account. The temperature of the optical benches was almost fully adjusted after less than twelve hours, \SI{90}{\percent} of the temperature shift happened within less than three hours. The RVs took more than 140 hours to adjust, most likely due to the zerodur spacer taking a longer time to adjust. This also means, that due to the inertia of the system a much larger step than \SI{7}{\milli\kelvin} is required to actually cause an RV shift of more than \SI{10}{\centi\meter\per\second}. 

\subsection{Temperature stability} 
\begin{figure}
    \centering
    \includegraphics[width=1\linewidth]{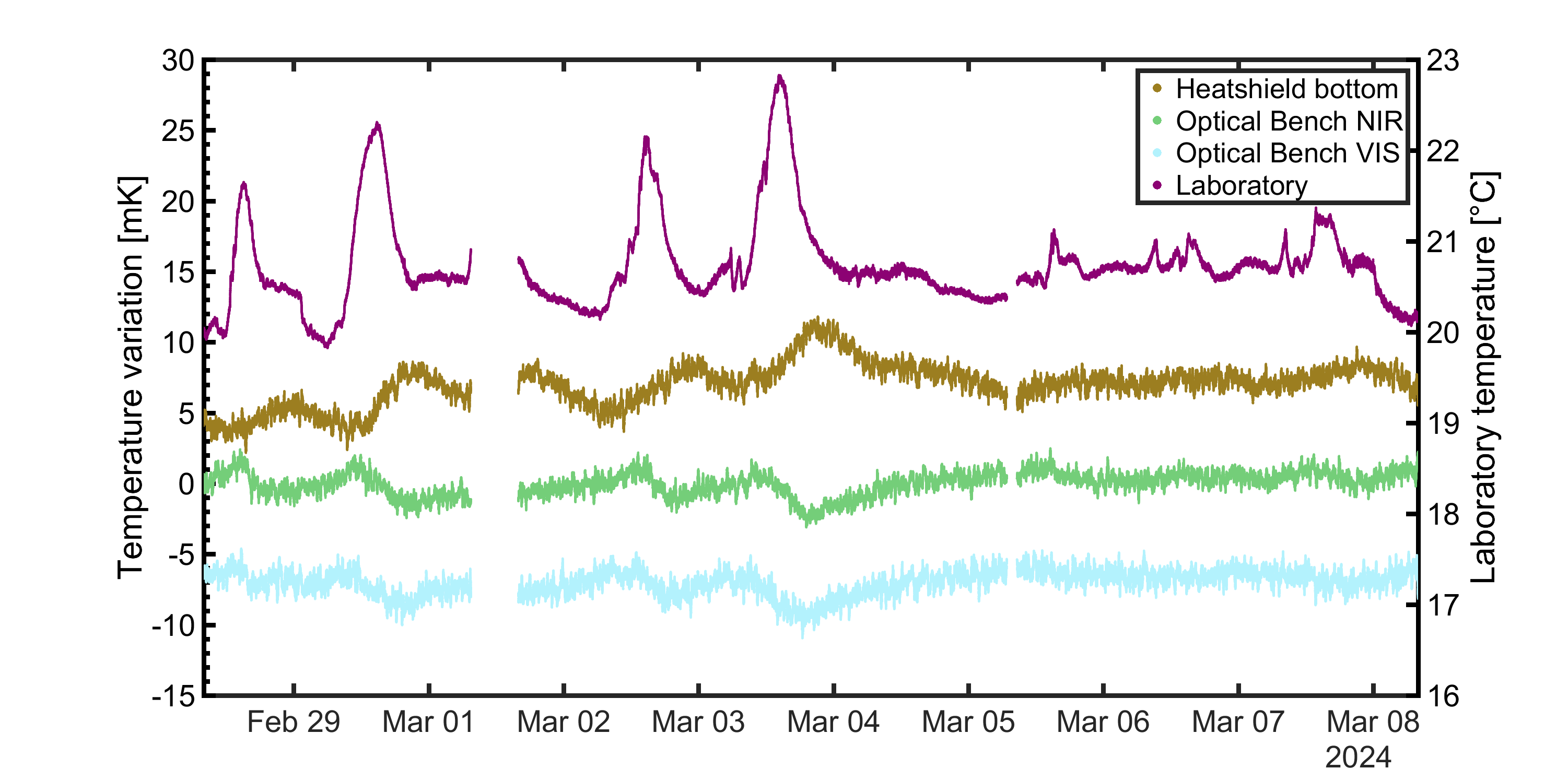}
    \caption{Measured temperatures at the VIS and NIR optical benches  and bottom heatshield temperature sensors (left axis) as well as laboratory temperature (right axis).}
    \label{fig:tempstable}
\end{figure}

After placing the vacuum vessel in its Styrodur enclosure the peak-to-valley temperature variation of the VIS optical bench was \SI{6.3}{\milli\kelvin} and that of the NIR optical bench was \SI{5.6}{\milli\kelvin} over a period of nine days (see Tab.\,\ref{tab:tempPtV}. Even the bottom heat shield showed a peak-to-valley temperature variation of only \SI{9.7}{\milli\kelvin} in this timeframe. The maximum and average daily peak-to-valley temperature variations of both optical benches are below \SI{6}{\milli\kelvin} and \SI{5}{\milli\kelvin}, respectively, so that even the more strict empirical temperature stability requirement derived above is fulfilled. The maximum daily variations corresponded to lab temperature variations of about \SI{2.4}{\kelvin}. The graph of the temperature measurements also shows that there are short timescale temperature variations (see Fig.\,\ref{fig:tempstable}). These probably neither affect the alignment of the setup due to the high thermal mass of the optical bench nor the length of the zerodur FP spacers due to the low thermal conductivity of the glass. When applying a moving mean of 30 minutes (a typical exposure time) the maximum daily temperature variations of the optical benches fall to a level of less than \SI{4}{\milli\kelvin} (average \SI{2.5}{\milli\kelvin}).

\begin{table}[htb]
\centering
\begin{tabular}{l c c c}
     Sensor &  Total [mK] & Daily max [mK] & Daily average [mK]   \\\hline
     Optical bench VIS & 6.3 & 5.9 & 4.3 \\
     Optical bench NIR & 5.6 & 4.8 & 3.8 \\
     Bottom heatshield & 9.7 & 6.3 & 4.8 \\
\end{tabular}
\caption{Measured peak to valley temperature variations of the optical benches and bottom heatshield temperature sensors over a nine day period when the setup was fully enclosed in styrodur insulation.}\label{tab:tempPtV}
\end{table}

\subsection{D-Curve}
\begin{figure}[htb]
\centering
 \includegraphics[width=0.9\textwidth]{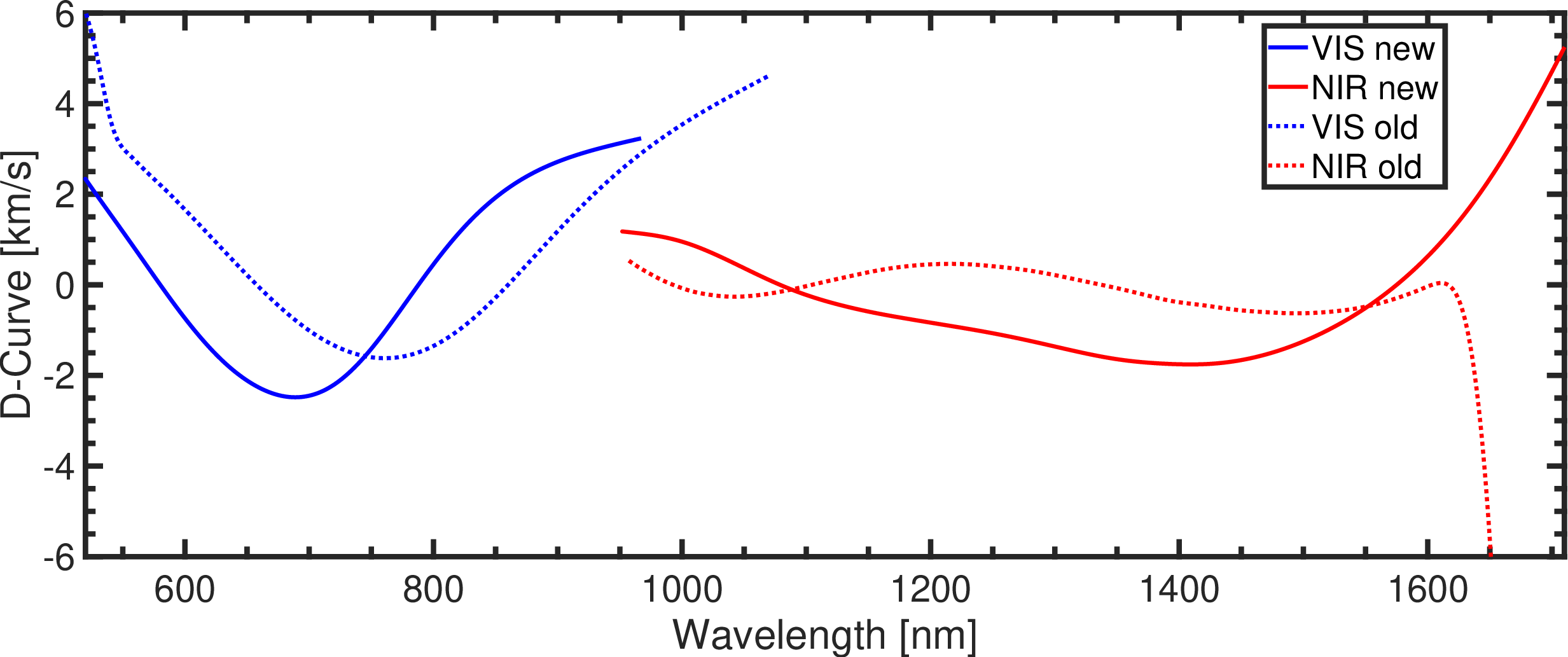}
 \caption{Comparison of the D-curves of the old and new etalons. The D-curves for the old etalons (dotted lines) were measured with CARMENES, the D-curves for the new etalons (solid lines) were measured with an FTS at IAG.} \label{fig:dcurve}
\end{figure}

For both etalons we recorded spectra with the light source that will also be used at the Calar Alto observatory to create master spectra. For the NIR FP  183 spectra, for the VIS FP 166 spectra were recorded. From these master spectra we can calculate the d-curve or effective cavity width \cite{Bauer2015} which is shown in Fig.\,\ref{fig:dcurve}. Comparing these with the d-curves of the old CARMENES FPs (calculated from CARMENES DATA) we can see a generally similar  behavior except in the regions were the coatings of the old FPs were out of the specified region. In these, the new FPs show a more well behaved trend.

\subsection{Radial velocity stability}
\begin{figure}[htb]
\centering
 \includegraphics[width=\textwidth]{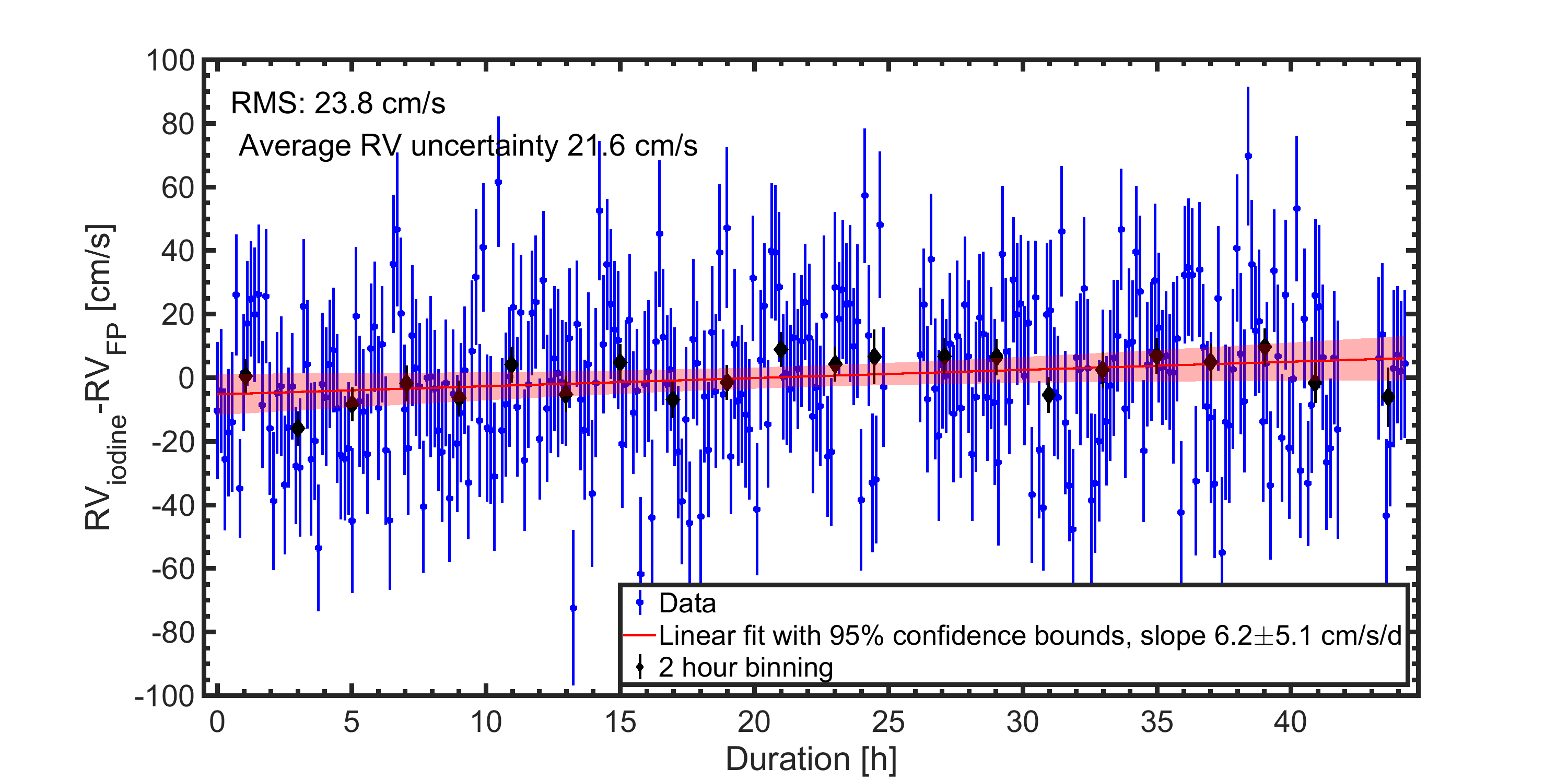}
 \caption{The RV drift of the FP compared against an iodine cell to take out the instrumental drift of the FTS (blue circles), a two hour binning of the data (black diamonds) and a linear fit (red line) are displayed as well.\vspace{-0.4cm}} \label{fig:rvs}
\end{figure}

\begin{figure}[htb]
\centering
 \includegraphics[width=\textwidth]{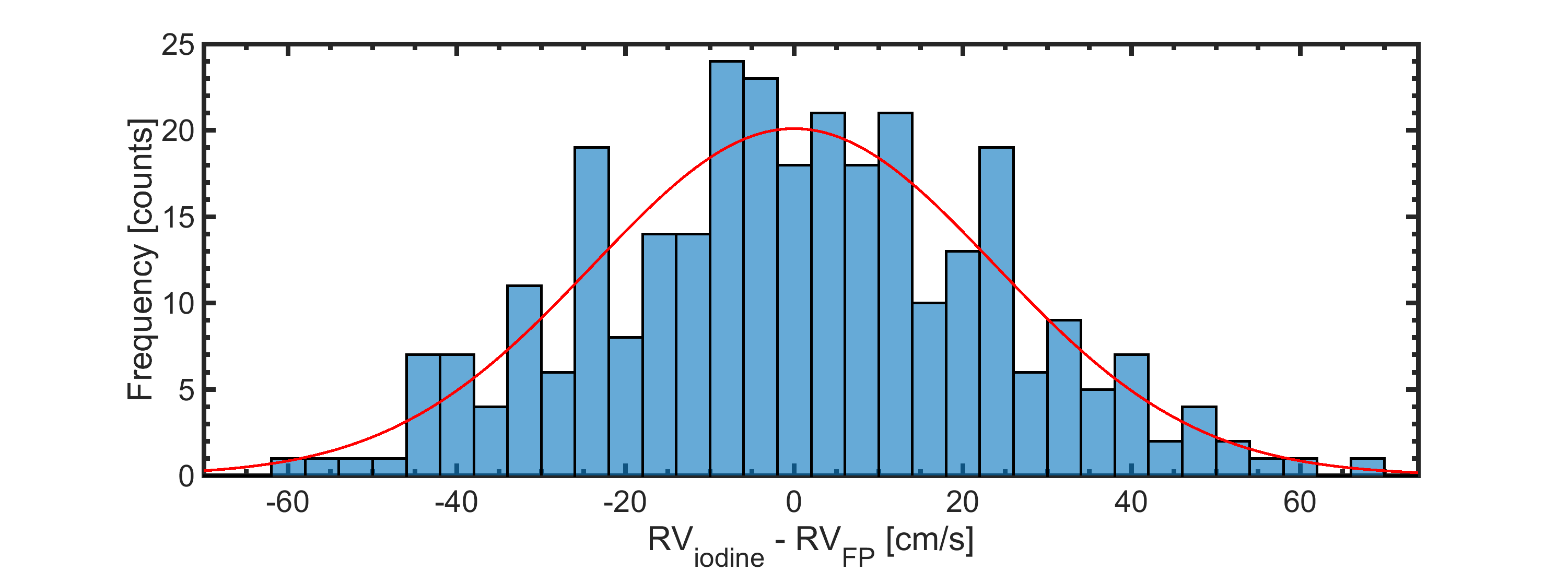}
 \caption{Histogram of the RV drift of the FP compared against an iodine cell to take out the instrumental drift of the FTS.} \label{fig:rvhist}
\end{figure}

To verify the RV stability of the FP we recorded 300 measurements over 44 hours with the FP illuminated by two LEDs (for better SNR) and an iodine cell in parallel. This is practically the same setup as in Debus et al.\cite{debus_towards_2023}, only with a different FP with fewer lines to account for the lower resolving power and thus also reduced precision. We calculated the RVs of both the FP and iodine using SERVAL.\cite{Zechmeister2017} The difference between FP and iodine is shown in Fig.\,\ref{fig:rvs}. The RV difference between both sources has an overall RMS of \SI{23.8}{\centi\meter\per\second} and an average intrinsic uncertainty \SI{21.6}{\centi\meter\per\second}. This indicates, that we are close to photon noise limited with only minimal systematics (see also Fig.\,\ref{fig:rvhist}). When binning the data over two hours, we get an
RMS of \SI{6.0}{\centi\meter\per\second} and an average uncertainty of \SI{5.7}{\centi\meter\per\second} per bin. There is a small linear trend of \SI{6.2(5.1)}{\centi\meter\per\second\per\day}
which is within the typical range of bulk RV drifts displayed by FPs such as ESPRESSO and HPF \cite{Terrien2021,Schmidt2022}. We therefore expect that the FP does not introduce any relevant drift or systematics into the CARMENES precision RV measurements.
 \vspace{-0.4cm}

%% file: sections/05-integration.tex
\section{Integration} \label{sec:integration} 
\begin{figure}[htb]
\centering
 \includegraphics[width=1\textwidth]{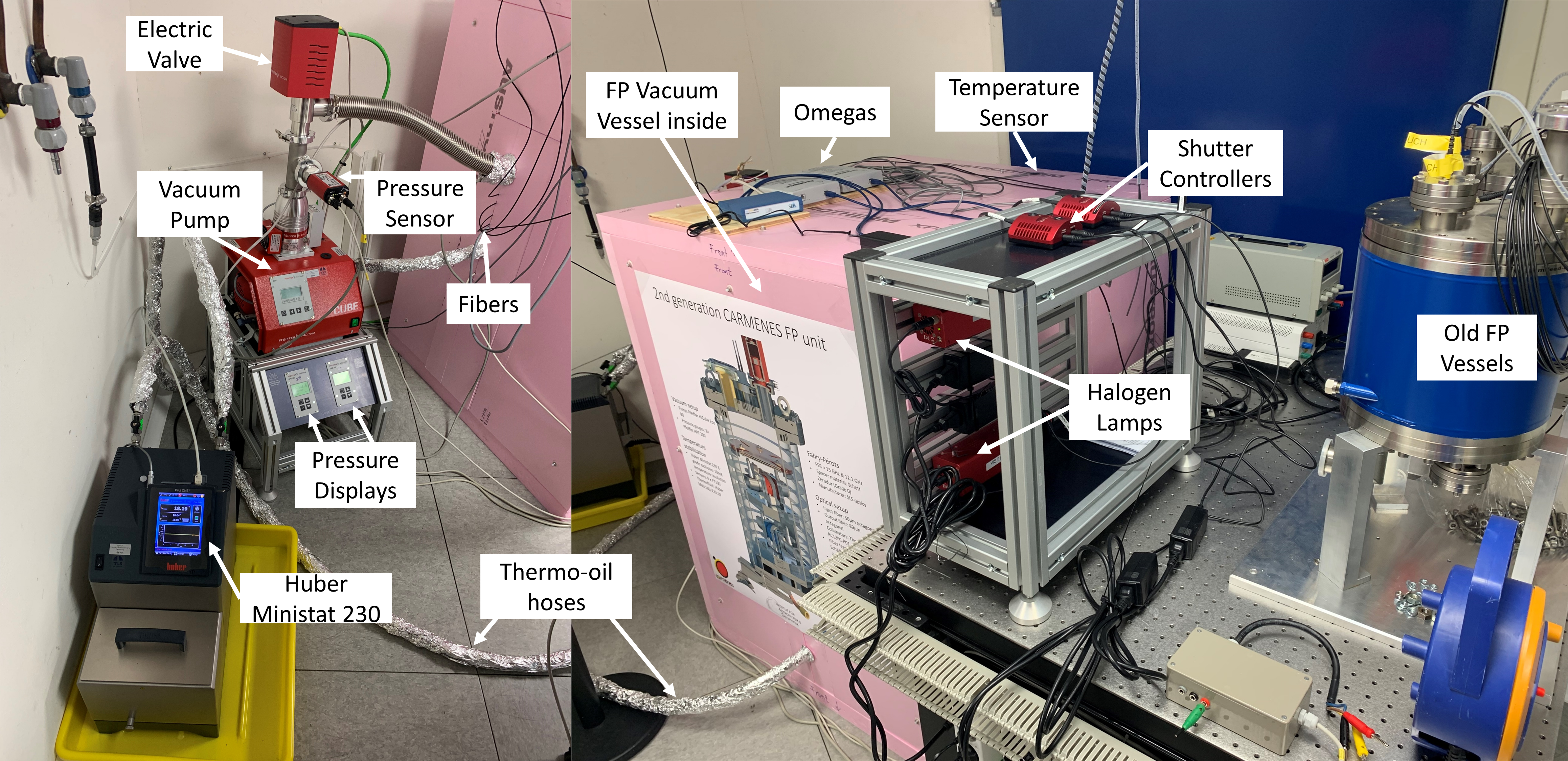}
 \caption{The new FP unit with all its peripheral in the calibration room of CARMENES.} \label{fig:caha_setup}
\end{figure}

In April 2024 the new FP unit has been successfully installed at the \SI{3.5}{\meter} telescope at Calar Alto Observatory, the home of the CARMENES instrument. It was a 1:1 replacement of the old two FP units with the vacuum pump (a Huber HiCube Eco 80) and the Huber Ministat 230 thermopump from the first generation FP being reused. Fig.\,\ref{fig:caha_setup}  gives an overview over the setup in the calibration room. The new \SI{89}{\micro\meter} output fibers are connected to the two calibration units (see Schäfer et al.\cite{Schaefer2018}) exactly as the old FP unit. In addition to the new FP unit with its styrodur enclosure a number of additional temperature sensors have been placed in the calibration room to further track the effect of the new insulation and temperature control.

\begin{figure}[htb]
\centering
 \includegraphics[width=\textwidth]{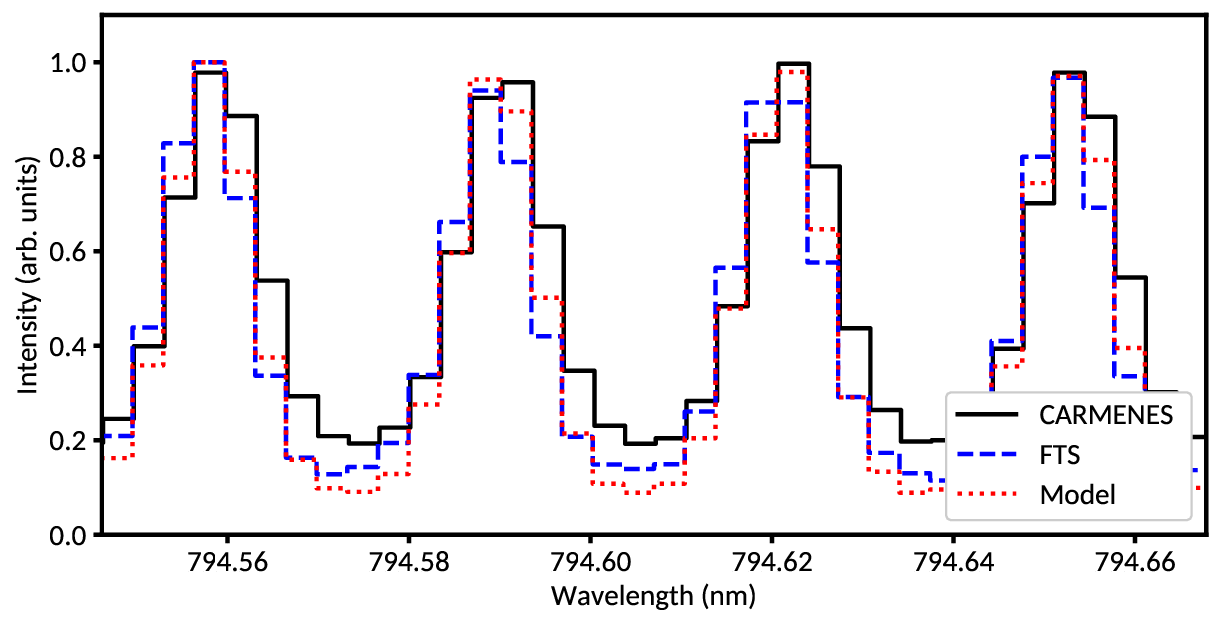}
 \caption{Comparison of the spectra shown in Fig.\, \ref{fig:vismodelvsfts} (FTS and forward simulation) with the new VIS FP spectrum as measured by CARMENES after the on-site integration at Calar Alto (black solid line).} \label{fig:vismodelvsftsvscar}
\end{figure}

\begin{figure}[htb]
\centering
 \includegraphics[width=\textwidth]{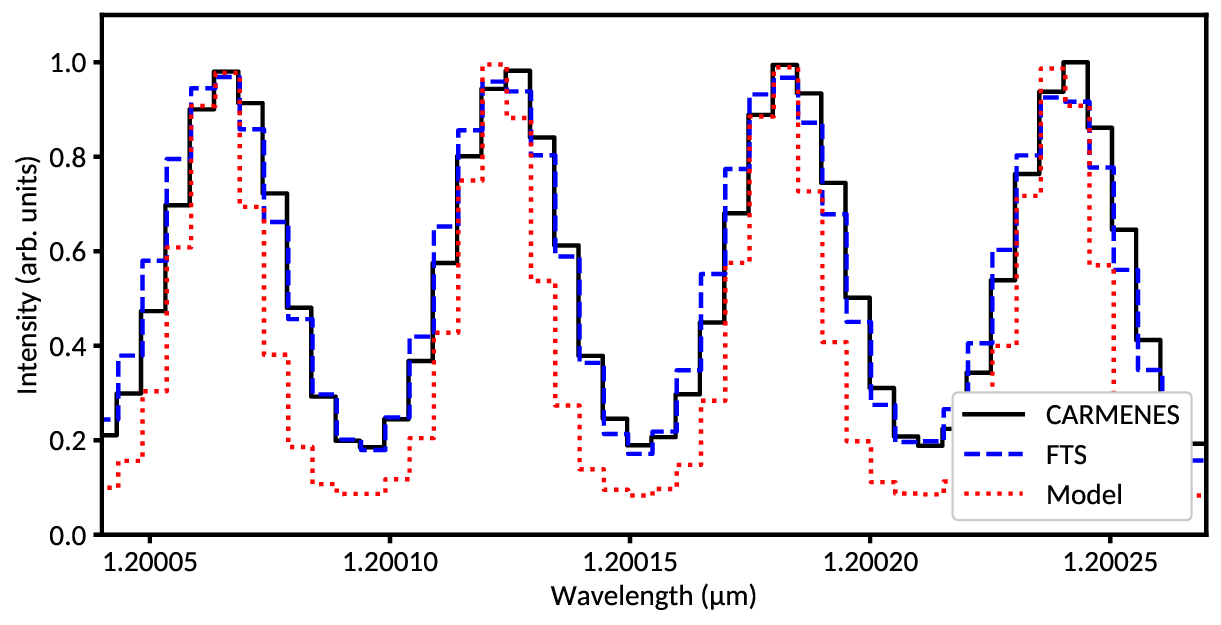}
 \caption{Comparison of the spectra shown in Fig.\,\ref{fig:nirmodelvsfts} (FTS and forward simulation) with the new NIR FP spectrum as measured by CARMENES after the on-site integration at Calar Alto (black solid line).} \label{fig:nirmodelvsftsvscar}
\end{figure}

Initial tests during integration using CARMENES show a big improvement in contrast compared to the old setup. The VIS FP shows slightly degraded performance compared to the lab measurements (see Figs.\,\ref{fig:vismodelvsftsvscar} and\,\ref{fig:nirmodelvsftsvscar}), probably caused by a slight misalignment of the optics during the transport: The contrast is reduced to 80.8\,\% while the FTS had shown 87.9\,\%. The NIR FP contrast is very close to the previous FTS measurements (81.4\,\% vs 83.9\,\%). 

While not perfect these values are a huge improvement to the first generation FP units  (compare Fig.\,\ref{fig:oldnewvismodelvscar}) and since the main focus of the upgrade is the improved temperature stability we decided to not re-open the vacuum vessel to repeat the alignment. 

The temperature stability could not yet be confirmed because the temperature measurements are not yet available for analysis. However, it can be expected to take at least 1-2\,weeks until the temperature gradient inside the system has been equalized. From the RV-drift measured during thermalization in our lab (see Fig.\,\ref{fig:rvsettling}) we can also expect the RVs to only become stable after more than a week, considering the much larger temperature step of \SI{10}{\celsius} from room temperature to control set point.

\FloatBarrier

%% file: sections/06-summary.tex
\section{Summary} \label{sec:conc}
We have successfully built and installed a new FP unit for the CARMENES spectrograph and verified its high RV precision with an FTS in our laboratory. We showed, that in a two hour binning an RV precision of \SI{6}{\centi\meter\per\second} is achieved. First measurements on sky should already be performed by the time of publishing. The initial outlook is promising, as the calibration data show an improved single order RV precision of the FP. With this new FP unit, the CARMENES spectrograph is well equipped for the coming CARMENES Legacy-Plus survey. This latest evolution of the concept first demonstrated by Schaefer et al. \cite{schafer_fabry-perot_2014} is a further step from the last iteration by Debus et al. \cite{debus_towards_2023} with improved temperature stability and the capability to install multiple FPs in a single vacuum vessel. 